# Topological light bullets supported by spatio-temporal gain

Valery E. Lobanov, Olga. V. Borovkova, Yaroslav V. Kartashov, Victor A. Vysloukh, and Lluis Torner

*ICFO-Institut de Ciencies Fotoniques, and Universitat Politecnica de Catalunya, Mediterranean Technology Park, 08860 Castelldefels (Barcelona), Spain*

We reveal that the competition between diffraction, cubic nonlinearity, two-photon absorption, and gain localized in both space and time results in arrest of collapse, suppression of azimuthal modulation instabilities for spatiotemporal wavepackets, and formation of stable three-dimensional light bullets. We show that Gaussian spatiotemporal gain landscapes support bright, fundamental light bullets, while gain landscapes featuring a ring-like spatial and a Gaussian temporal shapes may support stable vortex bullets carrying topological phase dislocations.

The generation of stable, fully three-dimensional, self-supported, long-lived light bullets is one of the grand challenges of the field of nonlinear optics that after several decades still remains open. They are spatiotemporal wavepackets that are localized in space and time in the presence of physical effects that otherwise cause spatial and temporal pulse-beam spreading and/or break-up [1]. Besides the fundamental importance of the matter, light bullets may be key building blocks in future ultrafast optical processing devices.

In conservative nonlinear media light bullets can exist due to the exact balance between nonlinearity and diffraction. However, despite the fact that they have been predicted to exist in various nonlinear materials [2,3], their stabilization represents a formidable challenge, especially in uniform media. In particular, spatiotemporal solitons suffer from supercritical collapse in conventional materials with cubic nonlinearity. Various approaches have been suggested to overcome this obstacle, including nonlinearity saturation [4], higher-order diffraction or dispersion [5], or nonlocality of the nonlinear response [6]. Notice that a different type of light bullets, in the form of Airy-Bessel wavepackets propagating in linear media, has been recently realized [7]. Nonlinear media with a shallow transverse modulation of refractive index (e.g., arrays of evanescently coupled waveguides) does support stable light bullets [8] whose signature over a few diffraction-dispersion lengths was observed by Minardi et.al., in a recent landmark experiment [9]. Longitudinally modulated waveguide arrays or tandem structures have been also suggested to observe long-lived light bullets [10]. Instabilities of three-dimensional solitons may be suppressed also in dissipative systems [11], where stabilization is achieved by the action of nonlinear dissipation and competing nonlinearities. Thus, stable dissipative light bullets were predicted in laser systems with saturable gain and absorption [12]. Stable fundamental light bullets [13] and three-dimensional vortex bullets [14] exist also in uniform dissipative systems governed by the complex cubic-quintic Ginzburg-Landau equation. The presence of a conservative external potential in such systems substantially enriches the properties and domains of existence of the spatiotemporal solitons [15]. Three-dimensional trapped structures exist also in optical cavities [16], although the field distribution in such states varies periodically or quasi-periodically in time in any spatial point in contrast to standard light bullets.

Very recently it was demonstrated that the evolution of nonlinear excitations in dissipative media with nonlinear losses is affected dramatically by a spatially modulated gain. Formation of solitons in inhomogeneous gain landscapes was studied in Bragg gratings and waveguides [17,18], in materi-als with periodic refractive index modulation [19], in Bose-Einstein condensates [20], and in uniform media with cubic nonlinearity and two-photon absorption [21,22]. In all such settings, solitons form in domains where gain is maximal, hence the topology of the gain landscape determines the soliton symmetry. While spatial localization of gain ensures the stability of the background, the nonlinear losses acting in the system may potentially suppress collapse, even in cubic nonlinear media. Such stabilization of solitons due to the interplay between localized gain and nonlinear losses has been shown in two-dimensional geometries. In this work we show that focusing cubic medium with gain localized in both space and time, and uniform nonlinear losses does support stable three-dimensional solitons. While spherically symmetric gain landscapes support stable fundamental solitons, the gain profiles featuring ring-like spatial and bell-like temporal shapes support stable vortex bullets.

We consider the propagation of a spatiotemporal wave packet along the $\xi$-axis in a cubic nonlinear medium with two-photon absorption and non-uniform gain that can be described by the nonlinear Schrödinger equation for the dimensionless light field amplitude $q$:

$$i\frac{\partial q}{\partial \xi} = -\frac{1}{2}\left(\frac{\partial^2 q}{\partial \eta^2} + \frac{\partial^2 q}{\partial \zeta^2}\right) - \frac{\beta}{2}\frac{\partial^2 q}{\partial \tau^2} - q|q|^2 + i\gamma(\eta,\zeta,\tau)q - i\alpha q|q|^2. \qquad (1)$$

Here $q = (2\pi L_{\text{diff}} n_2 / \lambda)^{1/2} E$; $E$ is the field amplitude; $n_2$ is the nonlinear Kerr coefficient; $L_{\text{diff}} = k_0 r_0^2$ is the diffraction length; $k_0 = 2\pi n_0 / \lambda$ is the wavenumber; $n_0$ is the unperturbed refractive index at the wavelength $\lambda$; $\eta = x/r_0$, $\zeta = y/r_0$, and $\xi = z/L_{\text{diff}}$ are the transverse and longitudinal coordinates normalized to the characteristic transverse scale $r_0$ and diffraction length $L_{\text{diff}}$, respectively; $\tau = t/t_0$ is the normalized (retarded) time, $\beta = L_{\text{diff}} / L_{\text{disp}}$, $L_{\text{disp}} = \left|\partial^2 k / \partial \omega^2\right|^{-1} t_0^2$ is the dis-persion length; $\gamma(\eta,\zeta,\tau) = \gamma_0(\eta,\zeta,\tau)L_{\text{diff}}/2$, where $\gamma_0(\eta,\zeta,\tau)$ is the linear gain that varies in space and time; and $\alpha = \alpha_0(2k_0n_2)^{-1}$ is the normalized coefficient of the two-photon absorption. In our numerical calculations we consider a medium with focusing nonlinearity and assume anomalous dispersion (i.e., we set $\beta=1$).

The key novelty that we set to explore is this paper is the gain localized in both space and time moving with the group velocity of the bullet. Suitable spatial shaping of the concentration of amplifying centers or spatially localized optical pumping in the material with focusing nonlinearity and anomalous group velocity dispersion can be used for the creation of localized spatial gain landscapes. Implementation of the running gain propagating with the group velocity of the spatiotemporal wave packet may be accomplished by using an adapted variant of the so-called tilted-pulse-front pumping approaches [23]. Such method was introduced in the beginning of the 80s as the group-velocity matching technique between the optical pump pulse and the generated/amplified radiation. By using such traveling-wave excitation of laser materials, especially dye solutions, semiconductors, and nonlinear crystals of the optical parametric amplifiers, extremely high gain and reduced amplified spontaneous emission could be obtained. Currently the tilted-pulse-front pumping is widely applied for the excitation of short-wavelength lasers, broadband frequency conversion, and high-field THz pulse generation by the optical rectification of femtosecond laser pulses.

Solitons in systems governed by Eq. (1) form due to a double balance: between localized gain and nonlinear losses, on the one hand, and between diffraction, dispersion, and focusing nonlinearity, on the other hand. Spatiotemporal localization of gain ensures stability of the background at $|\zeta|,|\eta|,|\tau| \to \infty$, which is a crucial ingredient for the stability of localized solutions. The formation of light bullets occurs around the maxima of gain landscape $\gamma(\eta,\zeta,\tau)$ where two-photon absorption compensates gain, thus preventing the uncontrolled growth of the optical field amplitude and sup-pressing collapse, even though the conservative nonlinearity is cubic. Since an exact balance between gain and losses is required in both space and time for three-dimensional soliton formation, the shape of soliton and its extent along $\eta, \zeta, \tau$ axes strongly depend on the shape of the amplifying domain, and on the strength of gain and nonlinear absorption. This follows from the integral condition of balance between gain and losses that can be derived from Eq. (1):

$$\alpha \iiint |q|^4 \, d\eta d\zeta d\tau = \iiint \gamma(\eta, \zeta, \tau) |q|^2 \, d\eta d\zeta d\tau. \tag{2}$$

First we address spherically symmetric spatiotemporal gain landscape described by the function $\gamma = p_i \exp(-\rho^2/d^2)$, where $\rho = (\zeta^2 + \eta^2 + \tau^2)^{1/2}$, $p_i$ is the gain parameter, and $d$ is the gain profile width. Further we set $d = 1.5$. However, we verified that the results reported here remain qualitatively similar for other values of gain profile width. The fundamental light bullets supported by such gain landscapes were searched in the form $q = w(\rho)\exp(ib\xi)$, where $b$ is the propagation constant, while the complex function $w = w_r + iw_i$ describes soliton shape. We found the soliton profiles numerically using a relaxation method taking into account the balance condition (2) and the fact that in dissipative systems the propagation constant and all other soliton parameters are determined by the gain and loss coefficients $p_i$ and $\alpha$. The typical shapes of simplest spherically-symmetric dissipative light bullets are shown in Figs. 1(a) and 1(b). Such solitons are always characterized by the nontrivial internal phase distributions (see also Fig. 3). There is a flow of energy from the domain with gain toward the periphery of material where only nonlinear losses are acting. Due to this energy flow light can strongly penetrate into domain with losses, especially in the low-energy limit. For fixed nonlinear losses $\alpha$ an increase of gain $p_i$ (that we further use as a control parameter in our system) is accompanied by the progressive soliton localization in the domain with gain. The internal phase chirp in soliton becomes more pronounced with increase of $p_i$ [compare Figs. 1(a) and 1(b)]. Notice that for sufficiently high $p_i$ values a narrow spike develops on top of broad pedestal in soliton shape [Fig. 1(b)].

The energy of a fundamental light bullet $U = \iiint |q|^2 \, d\eta d\zeta d\tau$ nontrivially depends on the gain parameter $p_i$ [Fig. 1(c)]. The initial decrease of $U$ is followed by growth, which is monotonic for $\alpha > 1$ and nonmonotonic for $\alpha < 1$. Light bullets can be obtained only when gain parameter exceeds a minimal value $p_i^{low}$. The derivative $dU/dp_i$ tends to infinity, although the energy value $U$ remains finite when $p_i \to p_i^{low}$. We found that $p_i^{low}$ grows almost linearly with the increase of nonlinear losses [Fig. 2(a)] and monotonically decreases with the increase of the width of domain with gain. Similarly to energy the propagation constant $b$ of light bullet is the nonmonotonic function of $p_i$ for small nonlinear losses, but already for $\alpha \sim 3$ the propagation constant monotonically increases with $p_i$ [Fig. 1(d)]. When $p_i \to p_i^{low}$ the propagation constant approaches zero.

To elucidate the stability of the light bullets we propagated them up to large distances $\xi \sim 10^4$ in the presence of added input noise, by solving the governing equations with a standard split-step fast Fourier algorithm. We found that for a fixed strength of two-photon absorption the fundamental light bullets are usually stable within a finite interval of gain coefficients $p_i^{low} < p_i < p_i^{upp}$ adjacent to minimal gain level $p_i^{low}$ required for soliton existence. With increase of nonlinear losses the domain of stability slightly broadens and shifts to higher values of gain [Fig. 2(a)]. The solitons belonging to stability domain rapidly clean up the noise and propagate in a stable fashion over indefinitely long distances (Fig. 3, left). Outside the stability domain at $p_i > p_i^{upp}$ spherically symmetric solitons decay or experience considerable shape transformations. Propagating perturbed spherically symmetric solutions, we surprisingly found that instability may result in their transformation into stable asymmetric light bullets featuring ellipsoid-like states (Fig. 3, center). Such asymmetric solitons have equal widths in two orthogonal axes (e.g., $\eta, \zeta$) and larger width along third axis (e.g., $\tau$), which is highly unexpected taking into account that all parameters in the system are either uniform or spherically symmetric. The orientation of longer axis of such ellipsoid-like bullets is arbitrary. Such asymmetric solitons are characterized by unusual asymmetric internal energy flows, in contrast to spherically symmetric states. Although the entire family of asymmetric solitons may be constructed by varying gain $p_i$, the stability domain for such states is usually very narrow and it is located near the upper border of stability domain for symmetric solitons. It should be stressed that stable fundamental light bullets can be also constructed when dispersion parameter $\beta \neq 1$. Such bullets always feature ellipsoid-like intensity distributions.

Next we address gain landscapes having ring-like spatial shapes and Gaussian temporal profiles, i.e. $\gamma = p_i \exp[-(r-r_c)^2/d^2]\exp(-\tau^2/g^2)$, where $r = (\eta^2 + \zeta^2)^{1/2}$, while $d, r_c$ are the width and radius of the amplifying ring, respectively, and $g$ characterizes gain duration. Here we set $r_c = 3.75$, $d = 1.5$, $g = 2.0$ and we also verified that the results reported remain qualitatively similar for other values of such parameters. We search three-dimensional vortex light bullets in the form $q = w(\eta, \zeta, \tau)\exp(im\varphi)\exp(ib\xi)$, where $\varphi$ is the azimuthal angle in the plane $(\eta, \zeta)$, $m$ is the topological charge. Since stable vortex light bullets are attractors of Eq. (1) their shapes can be obtained by direct propagation for large distances, $\xi \sim 10^4$, using as an input $q|_{\xi=0} = r^{|m|}\exp(-r^2 + im\varphi)\exp(-\tau^2)$. Under proper conditions, such pulses experience fast reshaping, emit radiation and asymptotically approach stationary vortex states (thus, spatially localized gain act-ing in combination with two-photon absorption suppresses not only supercritical collapse in three dimensions, but also azimuthal modulation instabilities that are destructive for vortices in the majority of focusing materials). Notice that this method guarantees the robustness of the output states and shows that they can be excited from different input conditions, as they are attractors of the system. Note that we tested stability of all obtained solutions by adding input noise. Intensity distribution of vortex light bullets looks like a torus [see Fig. 3(c) that illustrates stable propagation of such bullet in the presence of perturbation]. We generated stable light bullets with topological charges $m=1$ and $m=2$. Typical spatial field modulus and phase distributions in such states are shown in Fig. 4. The spatial radius of the resulting vortex bullets is always close to the radius of amplifying ring, although the field of light bullet may penetrate considerably into domain with losses at low $p_i$ values.

The existence domains of vortex light bullets were obtained by slowly varying gain parameter for each particular level of nonlinear losses. Therefore, the obtained domains correspond to the parameter range in which vortex bullets are stable (thus, they may be narrower than total existence domains for such solitons). We found that for moderate two-photon absorption stable vortex bullets exist for a limited interval of gain parameters $p_i^{\text{low}} < p_i < p_i^{\text{upp}}$ [Fig. 2(b)]. The domain of stability monotonically expands with $\alpha$ and both $p_i^{\text{low}}$, $p_i^{\text{upp}}$ grow with $\alpha$. Above upper border of stability domain vortex light bullets transform into steadily pulsating breather-like structures. At the same time, for sufficiently large values of nonlinear losses the upper border of stability domain was not detected and light bullets remain stable even at very high gain levels. When nonlinear losses are too small ($\alpha < 1$) we did not find stable states at all. The dependencies of propagation constant of vortex light bullet on gain parameter and its energy on propagation constant are shown in Figs. 5(a) and 5(b), respectively. Both $U(b)$ and $b(p_i)$ are monotonically growing functions. Notice the difference in $U(b)$ and $b(p_i)$ dependencies for solitons that feature upper stability border (e.g. at $\alpha = 2$) and solitons that are sta-ble even for high $p_\text{i}$ values (e.g. at $\alpha = 3$). It should be mentioned that vortex solitons with higher topological charge require considerably higher gain levels for their stabilization. Thus, at $\alpha = 3$ stable vortex bullet with $m = 1$ can be obtained already at $p_\text{i}^\text{low} \approx 1.6$, while vortex bullet with $m = 2$ requires for its stabilization the gain level $p_\text{i}^\text{low} \approx 7.3$.

Summarizing, we showed that the competition between localized gain and two-photon absorption in cubic nonlinear media may results in suppression of supercritical collapse and azimuthal modulation instabilities, and therefore in the formation of stable, self-supported fully three-dimensional light bullets. Depending on the gain profile bullets may exist as bright, fundamental solitons or as topological vortex solitons carrying nested phase dislocations.

The work of O.V. Borovkova was supported by the Ministry of Science and Innovation, Government of Spain, grant FIS2009-09928.

# Figure captions

Figure 1. (Color online) Profiles of spherically symmetric solitons at $p_i = 2.3$ (a) and $p_i = 6.8$ (b) for $\alpha = 2$. Energy (c) and propagation constant (d) versus gain parameter $p_i$. Circles correspond to solitons in panels (a) and (b). All quantities are plotted in arbitrary dimensionless units.

Figure 2. (Color online) Lower $p_i^{low}$ and upper $p_i^{upp}$ borders of stability domain for (a) fundamental light bullets and (b) vortex light bullets with $m = 1$. All quantities are plotted in arbitrary dimensionless units.

Figure 3. Isosurface plots showing intensity distributions at $\xi = 0$ (top) and $\xi = 1000$ (bottom) and illustrating stable propagation of fundamental light bullet at $p_i = 5.2$, $\alpha = 2$ (left), transformation of initial spherically symmetric fundamental bullet into asymmetric bullet at $p_i = 4.35$, $\alpha = 1.5$ (center), and stable propagation of vortex light bullet with $m = 1$ at $p_i = 2.8$, $\alpha = 3.5$ (right). Output phase distributions are shown in the middle row. In all cases the noise was added to input field distributions. All quantities are plotted in arbitrary dimensionless units.

Figure 4. (Color online) Field modulus (left) and phase (center) distributions at $\tau = 0$ as well as isosurface intensity plots (right) for stable vortex light bullets at (a) $m = 1$, $p_i = 2.8$,

$\alpha = 3.5$ and (b) $m = 2$, $p_i = 8.0$, $\alpha = 3.0$. All quantities are plotted in arbitrary dimensionless units.

Figure 5. (Color online) Propagation constant of vortex light bullet versus gain parameter $p_i$ (a) and energy of light bullet versus propagation constant (b). All quantities are plotted in arbitrary dimensionless units.

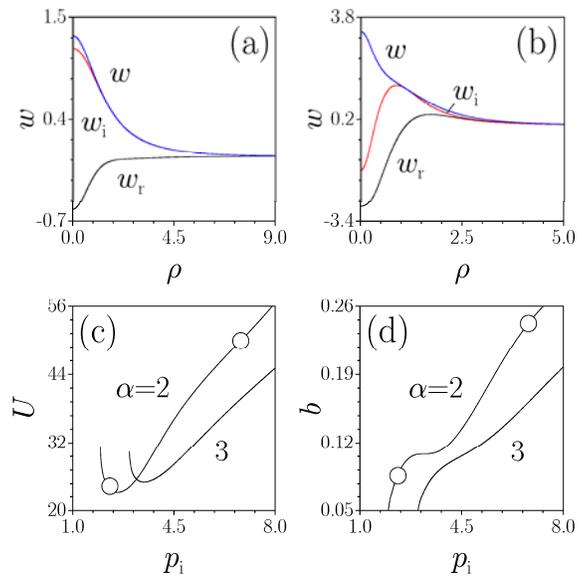

Figure 1. (Color online) Profiles of spherically symmetric solitons at $p_i = 2.3$ (a) and $p_i = 6.8$ (b) for $\alpha = 2$. Energy (c) and propagation constant (d) versus gain parameter $p_i$. Circles correspond to solitons in panels (a) and (b). All quantities are plotted in arbitrary dimensionless units.

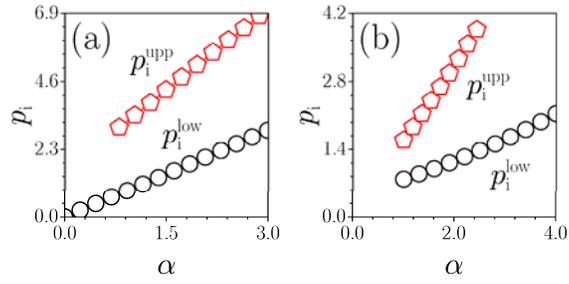

Figure 2. (Color online) Lower $p_i^{\text{low}}$ and upper $p_i^{\text{upp}}$ borders of stability domain for (a) fundamental light bullets and (b) vortex light bullets with $m=1$. All quantities are plotted in arbitrary dimensionless units.

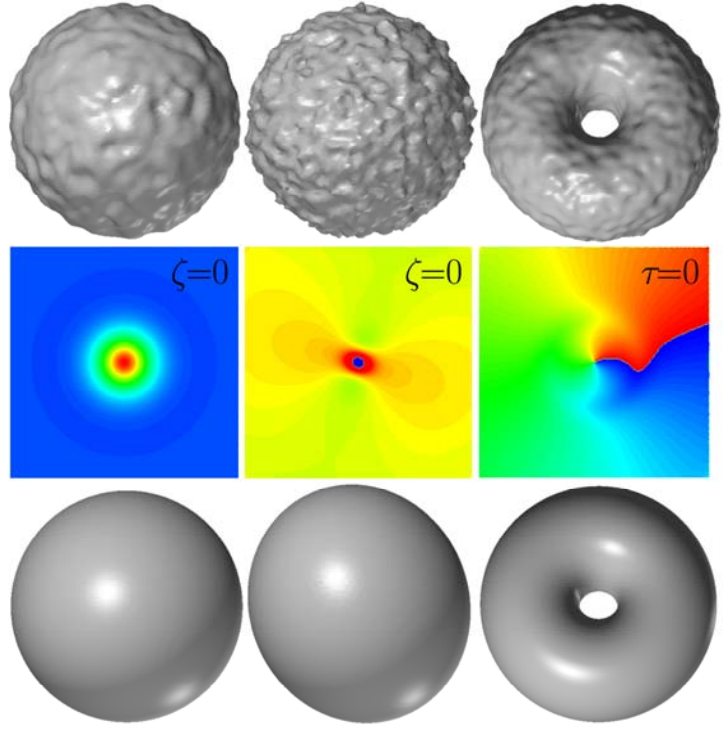

Figure 3. Isosurface plots showing intensity distributions at $\xi=0$ (top) and $\xi=1000$ (bottom) and illustrating stable propagation of fundamental light bullet at $p_i=5.2$, $\alpha=2$ (left), transformation of initial spherically symmetric fundamental bullet into asymmetric bullet at $p_i=4.35$, $\alpha=1.5$ (center), and stable propagation of vortex light bullet with $m=1$ at $p_i=2.8$, $\alpha=3.5$ (right). Output phase distributions are shown in the middle row. In all cases the noise was added to input field distributions. All quantities are plotted in arbitrary dimensionless units.

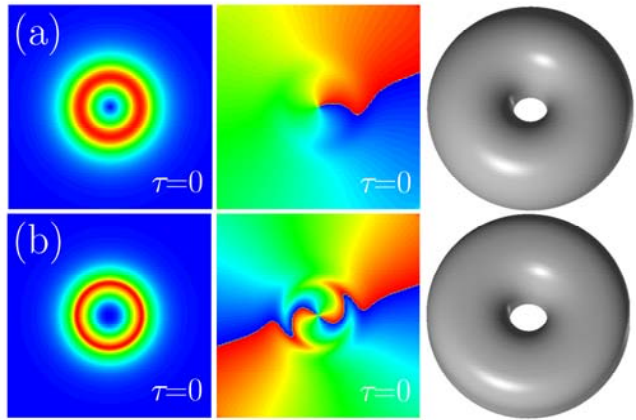

Figure 4.  (Color online) Field modulus (left) and phase (center) distributions at $\tau=0$ as well as isosurface intensity plots (right) for stable vortex light bullets at (a) $m=1$, $p_\mathrm{i}=2.8$, $\alpha=3.5$ and (b) $m=2$, $p_\mathrm{i}=8.0$, $\alpha=3.0$. All quantities are plotted in arbitrary dimensionless units.

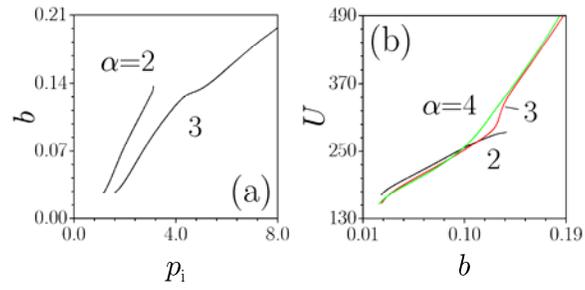

Figure 5. (Color online) Propagation constant of vortex light bullet versus gain parameter $p_{\mathrm{i}}$ (a) and energy of light bullet versus propagation constant (b). All quantities are plotted in arbitrary dimensionless units.